\documentclass[useAMS,usenatbib,usegraphicx,usebooktabs]{mn2e}
\usepackage{amssymb}
\usepackage{amsmath}
\usepackage{booktabs}
\usepackage{multirow}
\usepackage{hyperref}
\usepackage{times}

\title[The Fe\,K Line Profile of IRAS 18325-5926]
  {The Iron K Line Profile of IRAS 18325-5926}
\author[A.P. Lobban \& S. Vaughan]
  {A.P.~Lobban$^1$\thanks{e-mail: \href{mailto:al290@le.ac.uk}{al290@le.ac.uk}},
  S.~Vaughan$^1$ \\
  $^1$University of Leicester, X-Ray and Observational Astronomy Group, Department of Physics and Astronomy, Leicester, Leicestershire, LE1 7RH, U.K.}
\date{Accepted by MNRAS on 2 January 2014}

\pagerange{\pageref{firstpage}--\pageref{lastpage}} \pubyear{2014}

\def\LaTeX{L\kern-.36em\raise.3ex\hbox{a}\kern-.15em
    T\kern-.1667em\lower.7ex\hbox{E}\kern-.125emX}

\begin{document}

\label{firstpage}

\maketitle

\begin{abstract}
IRAS 18325-5926 is an X-ray bright, Compton-thin, type-2 Seyfert galaxy and it was the first Seyfert 2 in which the presence of a broad Fe\,K$\alpha$ emission line was claimed.  However, although the structure of the Fe line appears broad, there is tentative evidence that it may comprise multiple lines.  Nevertheless, previous analyses have only consisted of fitting stand-alone broad components to the Fe\,K band.  Here, we have analyzed all available X-ray CCD data from {\sl Suzaku}, {\sl XMM-Newton} and {\sl ASCA} to fully investigate the nature of the emission complex by testing broad-band physical models and alternative hypotheses.  We find that both a model consisting of broad, blurred reflection from an ionized accretion disc and a model consisting of cold, neutral reflection plus narrow emission lines from highly-ionized photoionized gas (log\,$\xi \sim 3.5$) offer statistically comparable fits to the data although the true reality of the Fe line cannot currently be determined with existing data.  However, it is hoped that better quality data and improved photon statistics in the Fe\,K band will allow a more robust distinction between models to be made.

\end{abstract}


\section{Introduction} \label{sec:Introduction}

The X-ray spectra of active galactic nuclei (AGN) are routinely observed to contain complex spectral features arising from both reprocessing and absorption of the primary continuum emission.  These features are often important diagnostics of the physical processes occurring within the immediate vicinity of an accreting supermassive black hole (SMBH).  Of particular interest is fluorescent line emission from near-neutral Fe\,K$\alpha$ at $\sim$6.4\,keV which appears to be a ubiquitous feature in type-1 Seyfert galaxies and is most likely produced via ``reflection'' from cold (near-neutral), optically-thick material such as the accretion disc and/or putative obscuring torus when illuminated by the central source (e.g. \citealt{NandraPounds94, Bianchi09, ShuYaqoobWang10, ShuYaqoobWang11, Patrick12}).  In many type-1 AGN the existence of a broadened Fe\,K$\alpha$ line has been claimed (e.g. \citealt{Nandra07}) with the observed broadening and associated strong red wing attributed to strong gravitational redshift a few $r_{\rm g}$\footnote{The gravitational radius is defined as $r_{\rm g} = GM / c^{2}$.} from the black hole (e.g. \citealt{Fabian89}).

To date, broad Fe\,K$\alpha$ emission lines have been detected in only a handful of obscured AGN; e.g. NGC 2992 \citep{Shu10} and NGC 5506 \citep{Guainazzi10}, both of which are type-1.8--1.9 Seyfert galaxies.  Additionally, similar features have been detected in the type-2 Seyferts MCG--5-23-16 \citep{DewanganGriffithsSchurch03} and IRAS 00521-7054, the latter of which is the first and only type-2 Seyfert galaxy in which the claim of the detection of black-hole spin has been made \citep{Tan12}.  The difficulty in detecting such broad features in type-2 AGN most likely arises from the presence of large absorbing columns in the line of sight.

The very first type-2 Seyfert galaxy in which a broad, skewed Fe\,K$\alpha$ line ($\sigma \sim 0.5$\,keV) was detected is IRAS 18325-5926 \citep{Iwasawa96} with a measured equivalent width as large as EW $\sim 600$\,eV.  Through observations with {\sl ASCA} and {\sl Chandra} the line has been observed to peak at $\sim$6.7--6.9\,keV (EW $=$ 200--600\,eV).  When coupled with the weak Compton-reflection hump $\gtrsim 10$\,keV, this suggests an origin in a highly-ionized accretion disc where the emission is dominated by Fe\,\textsc{xxv} with the observed broadening perhaps arising from Compton scattering and/or strong Doppler/gravitational effects due to its high inclination to the line of sight and proximity to the black hole.  Strong emission from Fe\,\textsc{xxv} has been thought to be uncommon in Seyfert galaxies and may have significant implications for the thermal structure of the disc \citep{NayakshinKazanasKallman00}.  Alternatively, highly-ionized Fe lines may possibly originate in distant photoionized gas, far from the disc (see \citealt{BianchiMatt02}).  Sources in which highly-ionized Fe lines have been detected include Mrk 205 \citep{Reeves01}, Mrk 766 \citep{Miller06} and NGC 7213 \citep{Bianchi08,Lobban10}.  The low-ionization nuclear emission-line region (LINER) galaxy M\,81 also displays two distinct emission lines from highly-ionized Fe at $\sim$6.7\,keV (Fe\,\textsc{xxv}) and $\sim$6.96\,keV (Fe\,\textsc{xxvi}) in addition to the near-neutral Fe\,K$\alpha$ core at $\sim$6.4\,keV \citep{Page04}.  Furthermore, more recent results with {\sl Suzaku} suggest that such highly-ionized features may be more common in AGN than previously thought with \citet{Patrick12} detecting narrow emission features from Fe\,\textsc{xxv} and Fe\,\textsc{xxvi} in 52 and 39\,per cent respectively of a sample of 46 nearby type-1 Seyferts.  Nevertheless, the detections of several of these lines in the same object are rare.

IRAS 18325-5926 is an X-ray-bright ($F_{\rm 2-10} \sim 1$--$3 \times 10^{-11}$\,erg\,cm$^{-2}$\,s$^{-1}$), Compton-thin type-2 Seyfert galaxy \citep{deGrijp85} at $z = 0.01982$ \citep{Iwasawa95} with an X-ray luminosity of $L_{\rm 2-10} \sim 10^{43}$\,erg\,s$^{-1}$ \citep{Iwasawa04}.  It displays peculiar spectral features which are unusual for type-2 Seyferts; namely a broad wing to the optical H$\alpha$ line (FWHM $\sim 3\,400$\,km\,s$^{-1}$) plus relatively strong low-ionization lines such as [O\,\textsc{i}]$\lambda$6\,300 and [N\,\textsc{i}]$\lambda$5\,199 \citep{Carter84} and a moderately low absorption column of $N_{\rm H} \sim 10^{22}$\,cm$^{-2}$.  The source also displays rapid X-ray variability (on occasion varying in flux by a factor of 2 on $10^{3}$--$10^{4}$\,s timescales; \citealt{Iwasawa95}) and the overall X-ray properties $> 2$\,keV are generally more akin to type-1 Seyferts.  The similarities between the X-ray and optical extinction led \citet{Iwasawa95} to suggest that the nucleus of IRAS 18325-5926 may be obscured globally as opposed to just by an optically-thick torus, as usually envisaged in the standard unification scheme for AGN \citep{Antonucci93}.  

Here we investigate the nature of the Fe-line profile in IRAS 18325-5926 by analysing all available CCD data.  We primarily focus on a 78\,ks observation with {\sl Suzaku} \citep{Mitsuda07}, the highest $S$/$N$ view of the Fe line of this object to date.  Additionally, including data acquired with {\sl Suzaku}'s hard X-ray detector (HXD) extends the observable bandpass up to $\sim$70\,keV allowing robust constraints on the continuum to be made while simultaneously allowing for the presence of any hard excess to be tested for which is important in pinning down the strength of any Compton-reflection component in this source.  In Section~\ref{sec:previous_observations}, we also consider data acquired from previous observations with {\sl XMM-Newton} and {\sl ASCA}.  As type-2 Seyferts may offer alternative insights into the nature of broad Fe lines, the true reality of such components should be assessed.  Therefore, our primary goal is to study the Fe line in the context of physical broad-band spectral models and test alternative hypotheses.

\section{Data analysis and reduction} \label{sec:data_analysis_and_reduction}

\subsection{{\sl Suzaku} data reduction}

IRAS 18325-5926 was observed by {\sl Suzaku} on 2007-10-26 for a duration of $\sim$170\,ks (sequence number: 702118010).  Here we discuss data acquired with the three operable X-ray imaging spectrometer (XIS; \citealt{Koyama07}) CCDs and the positive intrinsic negative (PIN) diodes of the HXD \citep{Takahashi07}.  There appears to be no other bright / contaminating sources within the XIS extraction region or the field of view (FOV) of the PIN.  Event files from version 2.3.12.15 of the {\sl Suzaku} pipeline processing were used and all data were reduced using HEA\textsc{soft} version 6.13 and {\sl Suzaku} CALDB version 20130110.

\subsubsection{XIS data reduction} \label{sec:XIS_data_reduction}

All XIS data were screened within \textsc{xselect} using standard selection criteria to exclude periods when the spacecraft passed through the South Atlantic Anomaly (SAA) plus the following 436\,s of such a passage.  Data acquired with Earth elevation angles (ELV) $\leq 5^{\circ}$ and Earth day-time elevation angles (DYE\_ELV) $\leq 20^{\circ}$ were also excluded.  Only good events with grades 0,\,2,\,3,\,4 and 6 were used while hot and flickering pixels were removed from the XIS images using the \textsc{cleansis} script.  Time intervals affected by telemetry saturation were also removed.

Source spectra from the XIS CCDs were extracted from circular regions of 3 arcmin radius centred on the source.  Background spectra were extracted from same-size regions but offset from the source region while avoiding the calibration sources on the corners of the CCD chips.  The data were acquired in the on-axis HXD-nominal pointing position.  XIS response files (RMFs) and ancillary response files (ARFs) were generated using the \textsc{xisrmfgen} and \textsc{xissimarfgen} \textsc{ftools} respectively including correction for the hydrocarbon contamination on the optical blocking filter \citep{Ishisaki07}.  A net XIS source exposure of 78\,ks was obtained for each of the three operable XIS chips.

The two front-illuminated XIS chips (XIS\,0 and 3; hereafter FI XIS\footnote{Note that XIS\,1 is back-illuminated and hereafter referred to as the BI XIS.}) were found to produce consistent spectra and so were combined to maximize signal-to-noise (S/N) using the \textsc{mathpha}, \textsc{addrmf} and \textsc{addarf} \textsc{ftools}.  The net source count rate for the two combined FI XIS CCDs was 1.062\,counts\,s$^{-1}$ in the 0.5--10\,keV energy range with the background rate $\sim$2.7 per cent of the source rate.  This count rate corresponds to an observed flux of $F_{\rm 0.5-10} = 2.98 \times 10^{-11}$\,erg\,cm$^{-2}$\,s$^{-1}$ ($F_{\rm 2-10} = 2.48 \times 10^{-11}$\,erg\,cm$^{-2}$\,s$^{-1}$)\footnote{Fluxes were derived from, and found to be consistent between, the best-fitting broad-band models described in Sections~\ref{sec:broad_fe-line_model} and~\ref{sec:narrow_fe_lines}.} and a luminosity of $L_{\rm 0.5-10} = 2.7 \times 10^{43}$\,erg\,s$^{-1}$ ($L_{\rm 2-10} = 2.2 \times 10^{43}$\,erg\,s$^{-1}$) assuming a distance of $\sim$86\,Mpc for this source.

\begin{figure}
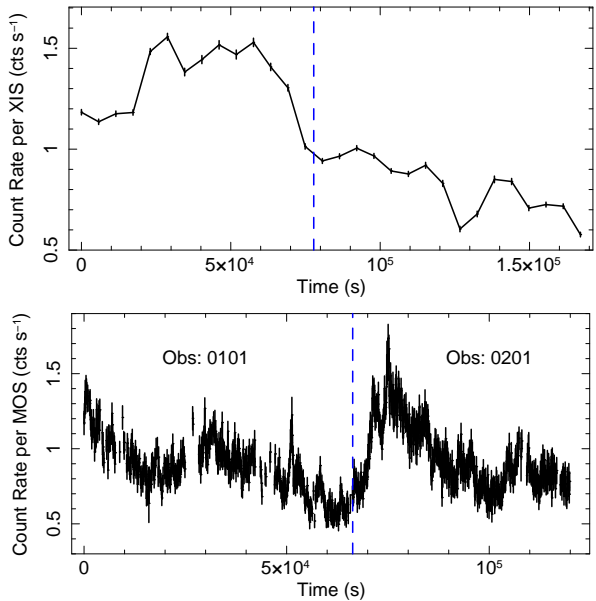

\begin{center}
\rotatebox{-90}{\includegraphics[width=4cm]{xis03_500-10000_5760s_perxis_split.ps}}
\rotatebox{-90}{\includegraphics[width=4cm]{mos12_200-10000_100s_20arcsec_permos_scaled.ps}}
\end{center}
\caption{Upper panel: The FI-XIS background-subtracted light curve of IRAS 18325-5926 in 5\,760\,s orbital bins from 0.5--10\,keV.  The vertical dashed line shows where the observation is split into high- and low-flux segments (see Section~\ref{sec:X-ray_spectral_variability}).  Lower panel: The {\sl XMM-Newton} MOS background-subtracted light curve in 100\,s bins from 0.5--10\,keV.  The short gaps arise from filtering out large background flares.}
\label{fig:lightcurves}
\end{figure}

Figure~\ref{fig:lightcurves} (upper panel) shows the 0.5--10\,keV {\sl Suzaku} FI-XIS light curve.  From a visual inspection it can be seen that the XIS count rate varies by a factor of $\sim$2 throughout the course of the observation\footnote{We note that the HXD-PIN light curve shows little evidence of any substantial short-term variability in the hard X-ray band.}.

\subsubsection{HXD data reduction} \label{sec:HXD_data_reduction}

As the brightness of IRAS 18325-5926 is below the detection threshold of the HXD gadolinium silicate (GSO) instrument we used data from the HXD PIN only which provides useful data over the 15--70\,keV energy range.  The source spectrum was extracted from a cleaned HXD PIN events file and screened to exclude periods when the spacecraft passed through the South Atlantic Anomaly plus the 180\,s before and the 500\,s after such a passage.  Data were also excluded with a cut-off rigidity (COR) $< 6$\,GeV/c and ELV $< 5^{\circ}$.  The instrumental background spectrum was generated from a `tuned' time-dependent model provided by the HXD instrument team \citep{Fukazawa09}.  Both the source and background spectra were made with identical good time intervals (GTIs) and the source exposure was corrected for the detector deadtime ($\approx 6.7$ per cent) using the \textsc{hxddtcor} script (version 2007 May).  A detailed description of the PIN detector deadtime is given in \citet{Kokubun07}.  After deadtime correction, the net exposure time of the PIN source spectrum was 69\,ks.  Note that the background spectral model was generated with 10 times the actual background count rate in order to minimize the photon noise on the background; this has been accounted for by increasing the effective exposure time of the background spectra by a factor of 10.  The HXD response file dated 2008-01-29 (epoch 4; HXD-nominal position) was used for all spectral fits involving the PIN.

To account for the cosmic X-ray background (CXB; \citealt{Boldt87}; \citealt{Gruber99}), a background spectrum was simulated which was generated from a model of the form $9.0 \times 10^{-9}(E/3\,{\rm keV})^{-0.29}{\rm exp}(-E/40\,{\rm keV})$\,erg\,cm$^{-2}$\,s$^{-1}$\,keV$^{-1}$.  This was normalized to the FOV of the HXD PIN instrument using a flat-field response and combined with the instrumental background file to produce a total background file.  The effective flux of the CXB component is $F_{\rm 15-50} = 7.80 \times 10^{-12}$\,erg\,cm$^{-2}$\,s$^{-1}$ which corresponds to a count rate of $\sim$0.017\,counts\,s$^{-1}$.  The net flux of IRAS 18325-5926 measured by the HXD over the same band is $1.74 \times 10^{-11}$\,erg\,cm$^{-2}$\,s$^{-1}$ and so the CXB component represents $\sim$45 per cent of the net source flux measured by the HXD PIN.  Note that there may be some uncertainty in the absolute flux level of the CXB component measured between missions; for instance, \citet{Churazov07} find the CXB normalization from {\sl INTEGRAL} to be about 10 per cent higher than measured by \citet{Gruber99} from the {\sl HEAO-1} data.  However, a factor of $\pm 10$ per cent uncertainty in the CXB normalization would result in a $\pm 4.5$ per cent uncertainty in the HXD flux for IRAS 18325-5926, which is well within the statistical uncertainty of the HXD PIN observations.  After background subtraction (including both the instrumental and CXB components), the resulting net PIN source count rate from 15--50\,keV is $0.043\pm0.003$\,counts\,s$^{-1}$ corresponding to a 15--50\,keV flux of $1.74\times10^{-11}$\,erg\,cm$^{-2}$\,s$^{-1}$.  Note that the total background count rate is $\sim$0.340\,counts\,s$^{-1}$ from 15--50\,keV with a typical $1 \sigma$ systematic uncertainty of $\pm 1.3$ per cent.

\subsection{{\sl XMM-Newton} data reduction} \label{sec:xmm-newton_data_reduction}

IRAS 18325-5926 was observed with {\sl XMM-Newton} on 2001-03-05 (ObsID: 0022940101) and 2001-03-06 (ObsID: 0022940201).  The EPIC cameras were operated in small-window mode using the medium filter.  However, due to technical problems, no useful data products were acquired with the EPIC-pn camera during either observation.  Nevertheless, the two EPIC-MOS cameras were operated successfully.  Version 13.0 of the Scientific Analysis Software (SAS) package was used for data reduction.  To correct for background flaring, data were filtered according to GTIs created by applying threshold count levels of $\leq 0.3$ and $\leq 0.2$\,counts\,s$^{-1}$ to the high-energy ($> 10$\,keV) MOS light curves for the first and second observations respectively.

To minimize background contribution, spectra were extracted from 20\,arcsec circular regions centered on the source.  Background spectra were extracted from much larger circular uniform regions away from both the central source and other nearby background sources.  The data were filtered for good X-ray events using the standard `FLAG==0' and `PATTERN $\leq$ 12' criteria.  This resulted in exposure times of 50\,ks and 49\,ks for the first and second observations respectively.  The averaged net source count rate for the two combined observations was 0.922\,counts\,s$^{-1}$ per MOS camera from 0.5--10\,keV with the background rate $<$1 per cent of the source rate.  This count rate corresponds to an observed flux of $F_{\rm 0.5-10} = 1.74 \times 10^{-11}$\,erg\,cm$^{-2}$\,s$^{-1}$ ($F_{\rm 2-10} = 1.47 \times 10^{-11}$\,erg\,cm$^{-2}$\,s$^{-1}$).

As the two EPIC MOS observations considered here are contiguous, both spectra from the MOS 1 camera were summed together, as were both spectra from the MOS 2 camera.  The two spectra were then fitted simultaneously, allowing the relative normalizations to go free.  Allowing other model parameters to go free between the two spectra always results in best-fitting parameters which are consitent to within 1 per cent.  We note that for plotting purposes, the spectra from both MOS cameras are combined to increase clarity.  The EPIC MOS lightcurve is shown in Figure~\ref{fig:lightcurves}.

\subsection{{\sl ASCA} data reduction} \label{sec:asca_data_reduction}

IRAS 18325-5926 was observed by {\sl ASCA} on 1993-09-11 and 1997-03-27.  The data have been described and analyzed previously by \citet{Iwasawa95,Iwasawa96,Iwasawa04}.  Here we briefly re-analyze data from the longer, more recent observation from 1997.  The {\sl ASCA} spectra were acquired directly from the Tartarus\footnote{http://heasarc.gsfc.nasa.gov/FTP/asca/data/tartarus/} database \citep{Turner01} with respective net exposure times for the solid-state imaging spectrometers (SIS) and gas imaging spectrometers (GIS) of 172 and 163\,ks respectively.  The averaged 0.5--10\,keV net source count rate was 0.363\,counts\,s$^{-1}$ for the two SIS instruments and 0.302\,counts\,s$^{-1}$ for the two GIS instruments with an observed flux of $F_{\rm 0.5-10} = 2.18 \times 10^{-11}$\,erg\,cm$^{-2}$\,s$^{-1}$ ($F_{\rm 2-10} = 1.86 \times 10^{-11}$\,erg\,cm$^{-2}$\,s$^{-1}$).  We note that all four spectra were fitted simultaneously but with an additional constant term in the model allowing the spectra to renormalize.

\section{Spectral analysis} \label{sec:Spectral_analysis}

The \textsc{xspec v12.8.0} software package \citep{Arnaud96} was used for spectral analysis of the background-subtracted spectra.  Data from 0.7--10\,keV were considered for use in both the FI-XIS and BI-XIS {\sl Suzaku} spectral fits although the 1.7--2.1\,keV band was ignored due to uncertainties in calibration associated with the instrumental Si\,K edge\footnote{http://heasarc.gsfc.nasa.gov/docs/suzaku/analysis/sical.html}.  In all fits the cross-normalization between the HXD-PIN and XIS detectors was accounted for by the addition of a fixed constant multiplicative component at a value of 1.18 for the HXD-nominal pointing position; a value derived using {\sl Suzaku} observations of the Crab (\citealt*{IshidaSuzukiSomeya07}\footnote{ftp://legacy.gsfc.nasa.gov/suzaku/doc/xrt/suzakumemo-2007-11.pdf}).  The relative normalization between the FI-XIS and BI-XIS spectra was allowed to vary, but best-fitting values were always within 1 per cent of each other.  The same was done for the two {\sl XMM-Newton} MOS cameras.  In all fits, the Galactic column density ($N^{\rm Gal}_{\rm H} = 6.47 \times 10^{20}$\,cm$^{-2}$) was also included (\citealt{DickeyLockman90}; \citealt{Kalberla05}) with the ``Tuebingen-Boulder'' (\textsc{tbabs}) code of \citet*{WilmsAllenMcCray00}.  Note that all fit parameters are given in the rest-frame of the galaxy.  A value for the Hubble constant of $H_{\rm 0} = 70$\,km\,s$^{-1}$\,Mpc$^{-1}$ was assumed and abundances are those of \citet*{WilmsAllenMcCray00} unless otherwise stated.

We begin by considering time-averaged {\sl Suzaku} spectra.  The XIS source spectra were binned to include $> 25$ counts per bin, thus enabling the use of $\chi^{2}$ minimization.  The HXD-PIN data were binned such that each bin has a detection significance of $\geq 5$\,$\sigma$.  Confidence intervals are quoted to 90 per cent confidence for one parameter of interest (i.e. $\Delta \chi^{2} = 2.706$) unless otherwise stated.

\subsection{The X-ray continuum} \label{sec:X-ray_continuum}

\begin{figure}
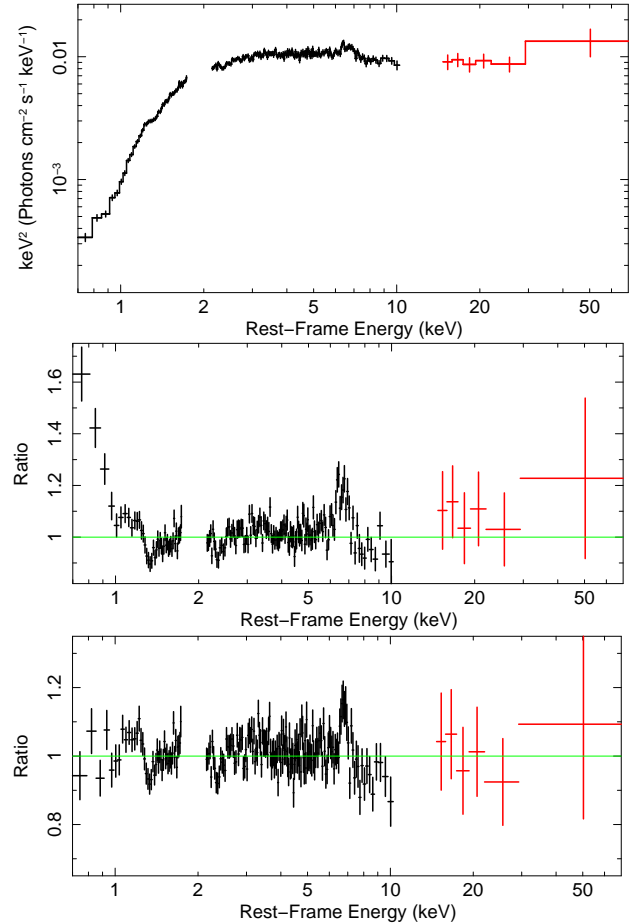

\begin{center}
\rotatebox{-90}{\includegraphics[width=4.5cm]{xis03_pin_eeuf.ps}}
\rotatebox{-90}{\includegraphics[width=3.85cm]{xis03_pin_tbabs_po_ratio.ps}}
\rotatebox{-90}{\includegraphics[width=3.85cm]{xis03_pin_tbabs_po_po_pexmon_ratio.ps}}
\end{center}
\caption{ Upper panel: The ``fluxed'' time-averaged {\sl Suzaku} X-ray spectrum of IRAS 18325-5926 from 0.7--50\,keV unfolded against a power law with $\Gamma = 0$.  Middle panel: The residuals of the time-averaged X-ray spectrum against a simple continuum fit of the form \textsc{tbabs}$^{\rm Gal}$ $\times$ \textsc{tbabs} $\times$ PL.  Lower panel: The residuals against a model of the form \textsc{tbabs}$^{\rm Gal}$ $\times$ (\textsc{tbabs} $\times$ (PL$_{\rm intrinsic}$ + \textsc{pexmon}) + PL$_{\rm scattered}$).  All FI-XIS data are shown in black while the HXD-PIN data are shown in red.  Data have been binned up and the BI-XIS data are excluded from the plots for clarity.}
\label{fig:xis03_pin_eeuf_and_model_residuals}
\end{figure}

\begin{figure}
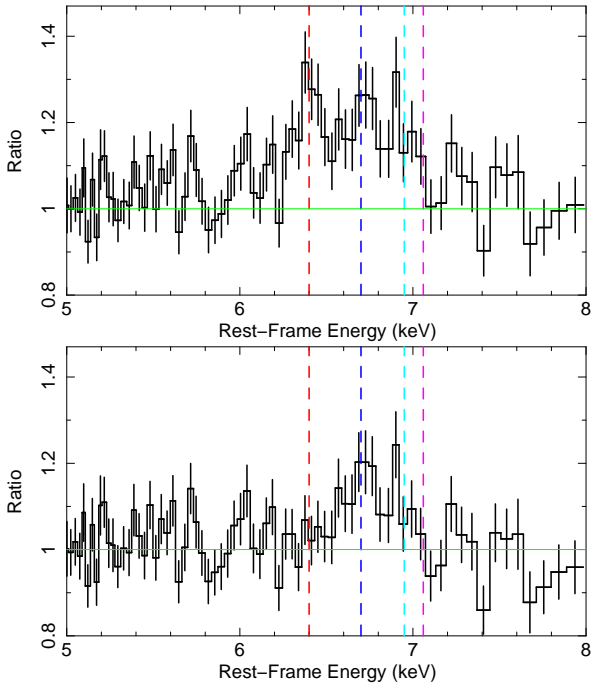

\begin{center}
\rotatebox{-90}{\includegraphics[width=4.5cm]{xis03_fe.ps}}
\rotatebox{-90}{\includegraphics[width=4.5cm]{xis03_fe_pexmon.ps}}
\end{center}
\caption{Upper panel: The ratio of the residuals of the FI-XIS data from 5--8\,keV compared to a power law absorbed by neutral material.  The vertical dotted lines show the expected line energies of, from left to right, Fe\,K$\alpha$ (red), Fe\,\textsc{xxv} $1s$--$2p$ resonance (navy), Fe\,\textsc{xxvi} $1s$--$2p$ Ly$\alpha$ (cyan) and Fe\,K$\beta$ (magenta).  Strictly speaking, the Fe\,\textsc{xxv}\,K$\alpha$ complex includes three other lines: two intercombination lines and a forbidden line, separated by $\Delta E \sim 45$\,eV, although these cannot be resolved with the current data.  Likewise the Fe\,\textsc{xxvi} line actually comprises a doublet peaking at 6.95 and 6.97\,keV.  Lower panel: The ratio of the residuals after the inclusion of the \textsc{pexmon} component to model the Fe\,K$\alpha$ emission at $\sim$6.4\,keV.  The spectra are binned up by a factor of 5 for clarity.}
\label{fig:xis03_fe}
\end{figure}

Figure~\ref{fig:xis03_pin_eeuf_and_model_residuals} (upper panel) shows the broad-band 0.7--50\,keV ``fluxed'' {\sl Suzaku} X-ray spectrum (see \citealt{Nowak05} and \citealt{Vaughan11}) of IRAS 18325-5926.  The observed counts spectrum has been previously published by \citet{Tripathi13}.  The most visually apparent feature is a significant column of absorbing material at energies $\lesssim 2$\,keV.  Fitting the continuum with a simple model consiting of a power law with a photon index of $\Gamma \sim 2.2$ modified by absorption from neutral material with a column density on the order of $N_{\rm H} \sim 10^{22}$\,cm$^{-2}$ ($\chi^{2}$/d.o.f. (degrees of freedom) = $1723/1430$) using the \textsc{tbabs} code reveals further spectral complexity, as shown in Figure~\ref{fig:xis03_pin_eeuf_and_model_residuals} (middle panel).  In particular, excess flux from 6--7\,keV is indicative of K-shell emission from Fe (also see Figure~\ref{fig:xis03_fe}; upper panel) and a soft excess can be seen at energies $\lesssim 1$\,keV.  Interestingly, the HXD-PIN data do not appear as a strong excess above the XIS data suggesting that any hard excess in this source may be relatively weak.

A model of the form \textsc{tbabs}$^{\rm Gal}$ $\times$ (\textsc{tbabs} $\times$ [PL$_{\rm intrinsic}$ + \textsc{pexmon}] + PL$_{\rm scattered}$) was applied to these data to include some of the fundamental components that one may expect in the X-ray spectrum of a Seyfert 2 (Figure~\ref{fig:xis03_pin_eeuf_and_model_residuals}; lower panel).  The \textsc{pexmon} model of \citet{Nandra07} is an additive component self-consistently incorporating the Compton-reflected continuum from a neutral slab combined with emission from Fe\,K$\alpha$, Fe\,K$\beta$, Ni\,K$\alpha$ and the Fe\,K$\alpha$ Compton shoulder.  Its purpose was to account for any hard excess of emission at energies $> 10$\,keV due to Compton scattering of intrinsic continuum photons while simultaneously modelling a narrow core of emission due to Fe\,K$\alpha$ fluorescence at $\sim$6.4\,keV (see Figure~\ref{fig:xis03_fe}; lower panel).  The photon index was tied to that of the primary power-law continuum ($\Gamma = 2.19 \pm 0.02$) and the abundances were fixed to solar values\footnote{Allowing the abundance of Fe ($A_{\rm Fe}$) to vary relative to other elements does improve the fit ($\Delta \chi^{2} = 10$); however, no tight constraint can be obtained on this parameter (i.e. $A_{\rm Fe} = 4.7^{+30.5}_{-2.9}$).} \citep{AndersGrevesse89}.  The inclination angle, $\theta_{\rm obs}$, was fixed at 60\,deg and allowing this parameter to vary does not improve the fit.  The cut-off energy of the power law was also fixed at a high value of $1\,000$\,keV; i.e. consistent with no cut-off in the observed bandpass.  Again, there is no requirement for this parameter to vary in the model fit.  The magnitude of the reflection scaling factor, $R$, is found to be $R = 0.36^{+0.11}_{-0.10}$, where a value of $R = 1$ corresponds to reflection from a semi-infinite slab subtending $2\pi$\,sr to the central source of X-rays\footnote{We note that the reflection scaling factor drops to $R = 0.18^{+0.08}_{-0.11}$ when $A_{\rm Fe}$ is allowed to vary.}.  The fit statistic achieved with the inclusion of the \textsc{pexmon} component is better by $\Delta \chi^{2} = 41$ (for one additional free parameter) when compared to the fit without it.

As noted by \citet{Tripathi13}, the shape of the soft excess appears to be continuum-dominated as opposed to arising from a series of narrow emission lines due to atomic transitions.  Consequently, the shape of the soft excess is largely reproduced by the secondary (`scattered') power-law component whose photon index is tied to that of the intrinsic power law ($\Gamma = 2.19 \pm 0.02$) but with their respective normalizations free to vary.  This is consistent with the findings of \citet{Tripathi13}.  Such a component could be representative of continuum emission that has been electron-scattered into the line-of-sight around an absorbing structure such as the torus.  The ratio of the normalizations of the power-law components then provides the ratio of scattered to primary flux which is $\sim$0.02, again consistent with \citet{Tripathi13}.  We note that there is no significant improvement to the fit upon allowing the photon index to vary between the two power-law components.

Further inspection of the soft-band residuals revealed a dip in the XIS spectrum at $\sim$1.3\,keV (see Figure~\ref{fig:xis03_pin_eeuf_and_model_residuals}; middle and lower panels).  Modelling this feature with a Gaussian with a negative normalization improves the fit by $\Delta \chi^{2} = 22$ and has best-fitting values of $E_{\rm c} = 1.33 \pm 0.02$\,keV and $\sigma < 50$\,eV with an equivalent width, EW $= 10 \pm 4$\,eV.  This absorption feature most likely corresponds to the $1s$--$2p$ transition from Mg\,\textsc{xi} (rest-frame energy $= 1.35$\,keV) and its blueshift would imply an origin in outflowing material with $v_{\rm out} \lesssim 9\,000$\,km\,s$^{-1}$.  The same feature was detected by \citet{Mocz11} in the {\sl Chandra} MEG spectrum and the best-fitting parameters are consistent\footnote{\citet{Mocz11} also find evidence for strong absorption features from Mg\,\textsc{xi} ($1s$--$3p$), Mg\,\textsc{xii}, Si\,\textsc{xiii} and Si\,\textsc{xiv} although there is no requirement for these features to be modelled in the {\sl Suzaku} XIS spectrum.  This may be due to the superior spectral resolution afforded by the {\sl Chandra} gratings.  Nevertheless, the upper limits on the equivalent widths of the Mg\,\textsc{xi} and Mg\,\textsc{xii} lines in the {\sl Suzaku} spectrum ($< 7$ and $< 6$\,eV respectively) remain consistent with the values obtained from the {\sl Chandra} gratings.  However, it should be noted that the transitions from Si fall inside the 1.7--2.1\,keV band that has been ignored in the XIS spectra and so their presence cannot be reliably tested for.}.

This simple model reproduces the general shape of the broad-band continuum well and results in a fit statistic of $\chi^{2}$/d.o.f. = $1564/1425$ where the bulk of the residuals are dominated by the Fe\,K band with excess emission apparent at $\sim$6.6--6.9\,keV (Figure~\ref{fig:xis03_fe}; lower panel).

\subsubsection{Application of the \textsc{mytorus} model} \label{sec:mytorus_model}

We also attempted to model the continuum with the \textsc{mytorus}\footnote{http://www.mytorus.com/} model of \citet{MurphyYaqoob09}.  The advantage of this model is that it accounts for both the absorption from cold material and the Compton-scattered `reflected' continuum in a self-consistent fashion and is valid in both Compton-thin and Compton-thick cases by providing continuous coverage between the two regimes.  As such, this is an ideal model to test with type-2 AGN such as IRAS 18325-5926.

The \textsc{mytorus} model assumes a toroidal geometry and comprises three components: the zeroth-order continuum (i.e. the directly transmitted component through the obscuring medium), the Compton-scattered continuum, and fluorescent emission lines including Fe\,K$\alpha$, Fe\,K$\beta$ and the associated Compton shoulder.  We replaced the \textsc{pexmon} component (from the model described in Section~\ref{sec:X-ray_continuum}) with the \textsc{mytorus} components such that the broad-band model took the form: \textsc{phabs}$^{\rm Gal}$ $\times$ ([PL$_{\rm intrinsic}$ $\times$ \textsc{myt}$_{\rm Z}$] $+$ [constant $\times$ \textsc{myt}$_{\rm S}$] $+$ [constant $\times$ \textsc{myt}$_{\rm L}$] $+$ Gauss$_{\rm abs}$ $+$ PL$_{\rm scattered}$).  Here, \textsc{myt}$_{\rm Z}$, \textsc{myt}$_{\rm S}$ and \textsc{myt}$_{\rm L}$ represent the zeroth-order continuum, the scattered continuum and the fluorescent line emission respectively.  Note also that the \textsc{tbabs}$^{\rm Gal}$ component has been replaced with the \textsc{phabs}$^{\rm Gal}$ component to allow for consistency with the abundances assumed by the \textsc{mytorus} model \citep{AndersGrevesse89}.  Gauss$_{\rm abs}$ is used to account for the Mg\,\textsc{xi} absorption line mentioned above.  

The photon index and normalization of the \textsc{mytorus} model were fixed to that of the power-law continuum, $z$ was fixed at the redshift of the source and the inclination angle and column density were tied between the various \textsc{mytorus} components.  This results in a fit statistic of $\chi^{2} / $d.o.f. $= 1574 / 1427$ with best-fitting values for the column density and inclination angle of $N_{\rm H} = 4.3^{+0.3}_{-1.9} \times 10^{22}$\,cm$^{-2}$ and $\theta_{\rm obs} = 63^{+3}_{-1}$\,deg. The parameters of the two power-law components remain unchanged with respect to the \textsc{pexmon} model described above.  However, this model under-predicts the amount of Fe emission at 6.4\,keV.  Allowing the relative normalizations of the \textsc{mytorus} components to vary significantly improves the fit.  By tying together the constants in front of the \textsc{myt}$_{\rm S}$ and \textsc{myt}$_{\rm L}$ components and allowing them to go free results in a much improved fit with $\chi^{2} / $d.o.f. $= 1535 / 1426$ with a best-fitting value for the free constant of $4.8^{+0.7}_{-1.0}$.  This adequately models the Fe-line peak at 6.4\,keV and is an improvement over the \textsc{pexmon} model described in Section~\ref{sec:X-ray_continuum} primarily due to a better fit of the HXD-PIN data $> 10$\,keV.  Allowing the relative normalizations to vary may account for time delays between the scattered and zeroth-order continua and could be representative of geometries, covering factors or elemental abundancies which deviate from the model assumptions.  We note that an equally good fit can be obtained by instead allowing the column density along the line of sight to be independent of that producing the scattered continuum.  In this instance, the line-of-sight column density becomes $N_{\rm H} = 3.8^{+0.6}_{-2.6} \times 10^{22}$\,cm$^{-2}$ whereas the column density of the scattering medium (which is tied to that of the material producing the fluorescent line emission) increases to $N_{\rm H} = 2.5^{+1.3}_{-0.8} \times 10^{23}$\,cm$^{-2}$.  This may be consistent with a scenario whereby our sightline grazes the edge of the torus ($\theta_{\rm obs} \sim 60$\,deg) while the bulk of the scattered continuum arises from denser clumps of material out of the line of sight.

\subsection{The Fe-line complex} \label{sec:fe-line_complex}

Figure~\ref{fig:xis03_fe} shows the residuals once the peak at $\sim$6.4\,keV has been accounted for with neutral reflection.  Several authors have attempted to model the Fe\,K emission from previous observations with a broad component (e.g. \citealt{Iwasawa96,Iwasawa04,Tripathi13}) assumed to originate in the accretion disc near to the black hole.  However, only simple Gaussian and relativistic disc-line profiles (e.g. \citealt{Fabian89}; \citealt{Laor91}) have been used to parametrize the {\sl Suzaku} Fe-line profile to date.  Additionally, \citet{Fukazawa11} analyzed a sample of 88 Seyfert galaxies observed with {\sl Suzaku} and accounted for the Fe\,K emission profile of IRAS 18325-5926 with two narrow Gaussians peaking at 6.4 and 6.7\,keV with respective equivalent widths of $38\pm11$ and $23\pm5$\,eV.  Here we take a broad-band approach to the {\sl Suzaku} data to fit more physical models while also testing alternatives.

\subsubsection{Broad Fe-line model} \label{sec:broad_fe-line_model}

\citet{Iwasawa04} suggested that the Fe\,K features could result from reflection off an ionized accretion disc.  A simple parametrization of the Fe\,K emission profile with a single broad Gaussian yields a best-fitting value for the centroid energy of $E_{\rm c} = 6.60 \pm 0.07$\,keV, consistent with an origin in ionized material while the best-fitting value for the intrinsic width is found to be $\sigma = 330^{+90}_{-70}$\,eV and the equivalent width is found to be EW $= 170 \pm 40$\,eV.  Some authors have also previously attempted to fit the profile with the \textsc{diskline} component of \citet{Fabian89} (e.g. \citealt{Iwasawa04}; \citealt{Tripathi13}). The variable parameters of the \textsc{diskline} model are the centroid energy, the inner and outer radii of emission in units of $r_{\rm g}$, the inclination angle of the source to the observer's sightline, $\theta$, the emissivity index, $q$, which defines the radial dependence of emissivity across the disc ($r^{-q}$) and the integrated intensity of the line. Parametrizing the Fe\,K profile this way requires a best-fitting rest-frame value for the centroid energy of $E_{\rm c} = 6.61^{+0.04}_{-0.05}$\,keV, again consistent with ionized Fe, with a measured equivalent width of EW $= 190^{+100}_{-80}$\,eV (line flux, $F = 4.2^{+1.1}_{-1.0} \times 10^{-5}$\,photons\,cm$^{-2}$\,s$^{-1}$). An upper limit on the inner radius of emission is found to be $r_{\rm in} < 80$\,$r_{\rm g}$ although the outer radius is unconstrained and so fixed at a high value of $r_{\rm out} = 500$\,$r_{\rm g}$. Additionally, the emissivity index and inclincation angle are constrained to be $q = 1.6^{+1.1}_{-0.6}$ and $\theta = 44^{+20}_{-10}$\,deg. The fit statistic achieved with this fit is $\chi^{2} / {\rm d.o.f.} = 281 / 223$ from 5.5--7.5 keV.

Here, to take a self-consistent broad-band approach, the model described in Section~\ref{sec:X-ray_continuum} was taken and, in addition to the \textsc{pexmon} component which models the narrow core of Fe\,K$\alpha$ emission from distant material at $\sim$6.4\,keV plus (weak) hard excess, the remaining, broader residuals were modelled with a relativistically-blurred reflection model computed using the \textsc{reflionx} code of \citet{RossFabian05} convolved with the \textsc{rdblur} code.  \textsc{rdblur} is a convolution kernel based on the \textsc{diskline} model of \citet{Fabian89} taking into account relativistic effects from an accretion disc around a Schwarzschild black hole.  The \textsc{reflionx} code is designed to model the emergent spectrum when a power law irradiates a photoionized, optically-thick slab of gas. The model assumes a high-energy exponential cut-off, $E_{\rm cut} = 300$\,keV, and uses the abundances of \citet{AndersEbihara82}. The advantage of the \textsc{reflionx} model is that it self-consistently models both the reflected continuum and line emission for many of the most important astrophysically-abundant ions.

The model parameters are the photon index ($\Gamma$) of the illuminating power law, which was tied to the photon index of the primary power-law continuum, the redshift ($z$), which was fixed at the redshift of the source, the abundance of Fe relative to the solar value (which was fixed at the solar value), the normalization of the reflected spectrum and the ionization parameter, which is defined as $\xi = L_{\rm ion} / nR^{2}$ and has units erg\,cm\,s$^{-1}$ where $L_{\rm ion}$ is the ionizing luminosity from 1--1\,000\,Rydberg in units erg\,s$^{-1}$, $n$ is the gas density in cm$^{-3}$ and $R$ is the radius of the absorbing / emitting material from the central source of X-ray in units cm.  The variable parameters of the \textsc{rdblur} code are based on those found in the \textsc{diskline} model described above; i.e. $r_{\rm in}$, $r_{\rm out}$, $\theta$ and $q$.  Allowing the parameters of \textsc{rdblur} to vary, we find the data are unable to meaningfully constrain them in the broad-band model.  Therefore, the inner and outer radii of emission were fixed at 6\,$r_{\rm g}$ and 500\,$r_{\rm g}$ respectively, the inclination angle was fixed at 60\,deg (consistent with the \textsc{pexmon} component) and the emissivity index was fixed at $q = 2$.

The best-fitting value of the ionization parameter of the reflector is $\xi = 1.0^{+1.1}_{-0.3} \times 10^{3}$\,erg\,cm\,s$^{-1}$, consistent with emission from an ionized disc.  The best-fitting value of the reflection scaling factor of the \textsc{pexmon} component then becomes $R = 0.15^{+0.11}_{-0.12}$.  These values are summarized in Table~\ref{tab:model_parameters}.  Finally, we note that further inspection of the residuals in the Fe\,K band reveals an absorption trough $> 7$\,keV.  A sweep across the 7--8\,keV band with the \textsc{edge} model within \textsc{xspec} results in best-fitting parameters for the energy of $E_{\rm edge} = 7.16^{+0.11}_{-0.05}$\,keV (consistent with \citealt{Fukazawa11}) and the optical depth of $\tau = 0.13 \pm 0.04$.  This improves the fit statistic by $\Delta \chi^{2} = 26$.

\begin{figure}
\begin{center}
\rotatebox{-90}{\includegraphics[width=6cm]{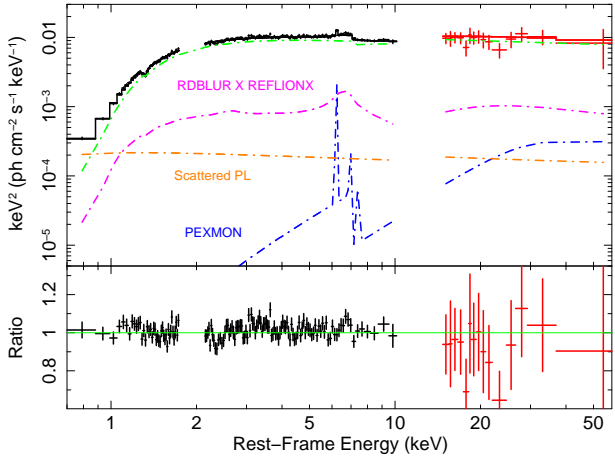}}
\end{center}
\caption{The fit to the broad-band {\sl Suzaku} spectrum of IRAS 18325-5926 when modelled with a relativistically-blurred reflection model, as described in Section~\ref{sec:broad_fe-line_model}.  The XIS and HXD-PIN data are shown in black and red respectively.  The additive model components are the primary absorbed power law (green), the scattered power law (orange), a neutral reflector (navy) and a blurred ionized reflector (magenta).  The bottom panel shows the data-to-model residuals.}
\label{fig:xis03_pin_tbabs_po_po_pexmon_rdblur_reflionx_gauss_edge_eeuf_ratio}
\end{figure}

The inclusion of these components results in an overall fit statistic of $\chi^{2}/$d.o.f. $= 1513 / 1421$.  We note that in addition to the standard fit statistic quoted over the full 0.7--50\,keV band, we also quote the fit statistic achieved just in the 5.5--7.5\,keV band which is $\chi^{2}$/d.o.f. $= 291 / 223$.  This is to avoid the statistic being dominated by the continuum fit and instead focuses solely on the residuals in the Fe\,K band.  This fit to the data is shown in Figure~\ref{fig:xis03_pin_tbabs_po_po_pexmon_rdblur_reflionx_gauss_edge_eeuf_ratio}.

\subsubsection{Modelling with narrow Fe lines} \label{sec:narrow_fe_lines}

As an alternative to reflection from the inner accretion disc, the emission profile may be modelled with individual peaks at $\sim$6.4, $\sim$6.7 and $\sim$6.9\,keV.  To test the plausibility of this model, the \textsc{pexmon} and \textsc{rdblur} $\times$ \textsc{reflionx} components were removed and three narrow Gaussians were added to model the three peaks.  The most significant peak can be fitted with a Gaussian with a best-fitting (rest-frame) centroid energy, $E_{\rm c} = 6.41 \pm 0.03$\,keV, and is found to be intrinsically narrow with $\sigma < 0.13$\,keV.  It has a line flux of $F_{\rm line} = 1.6^{+0.6}_{-0.4} \times 10^{-5}$\,photons\,cm$^{-2}$\,s$^{-1}$ and an equivalent width of EW $= 57^{+26}_{-25}$\,eV.  Modelling this feature improves the fit by $\Delta \chi^{2} = 60$ and would most likely correspond to K$\alpha$ emission from near-neutral Fe (i.e. Fe\,\textsc{i-xvii}).

\begin{figure}
\begin{center}
\rotatebox{-90}{\includegraphics[width=6cm]{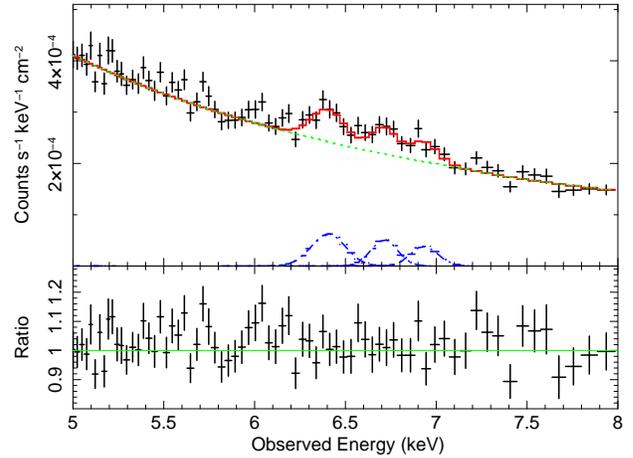}}
\end{center}
\caption{Upper panel: The best-fitting broad-band continuum model plus three narrow Gaussians (navy) fitted to the FI-XIS data from 5--8\,keV.  Lower panel: The ratio of the residuals to this model.}
\label{fig:xis03_fe_three_gaussian_data_ratio}
\end{figure}

\begin{figure}
\begin{center}
\rotatebox{0}{\includegraphics[width=8.5cm]{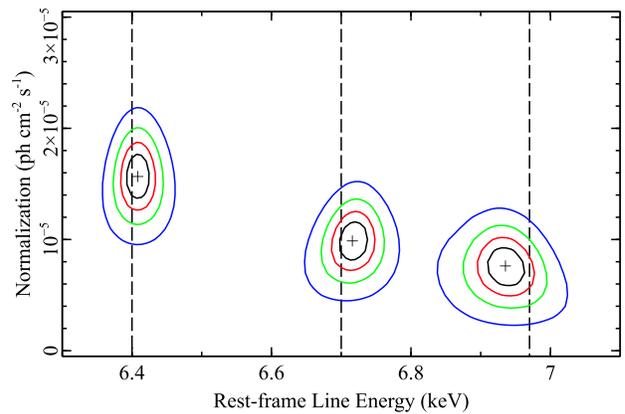}}
\end{center}
\caption{Two-dimensional contour plots between line energy and line flux generated across the Fe\,K band for the three narrow Gaussian peaks described in Section~\ref{sec:narrow_fe_lines}.  The inner contours correspond to 68\,per cent confidence and, increasing outwards, correspond to 90, 99 and 99.9\,per cent confidence respectively.  The vertical dashed lines show the rest-frame energies of emission from near-neutral Fe\,K$\alpha$ (6.40\,keV), Fe\,\textsc{xxv} $1s$--$2p$ resonance (6.70\,keV) and Fe\,\textsc{xxvi} Ly$\alpha$ (6.95--6.97\,keV).}
\label{fig:xis03_fe_three_gaussian_contours}
\end{figure}

The second and third peaks were both fixed to be intrinsically narrow ($\sigma = 10$\,eV).  The second peak has a best-fitting centroid energy of $E_{\rm c} = 6.72 \pm 0.04$\,keV and a line flux of $F_{\rm line} = 9.9^{+2.9}_{-3.0} \times 10^{-6}$\,photons\,cm$^{-2}$\,s$^{-1}$.  It has an equivalent with of EW $= 38^{+19}_{-13}$\,eV and is significant with $\Delta \chi^{2} = 26$.  The third peak requires a best-fitting centroid energy of $E_{\rm c} = 6.94^{+0.04}_{-0.05}$\,keV and a line flux of $F_{\rm line} = 7.6^{+2.8}_{-2.9} \times 10^{-6}$\,photons\,cm$^{-2}$\,s$^{-1}$.  It has an equivalent with of EW $= 33^{+18}_{-15}$\,eV and improves the fit by $\Delta \chi^{2} = 19$.  The two peaks may correspond to emission from the $1s$--$2p$ resonance transition of Fe\,\textsc{xxv} (rest-frame energy: 6.70\,keV; although possibly blended with additional unresolved peaks arising from the Fe\,\textsc{xxv} intercombination and forbidden lines) and blended emission from the Fe\,\textsc{xxvi}\,Ly$\alpha$ doublet (rest-frame energies: 6.95 and 6.97\,keV) respectively.  Figure~\ref{fig:xis03_fe_three_gaussian_data_ratio} shows this fit to the FI-XIS spectrum which corresponds to a fit statistic of $\chi^{2}/$d.o.f. $= 273 / 221$ from 5.5--7.5\,keV (a slight improvement over the corresponding \textsc{diskline} parametrization in Section~\ref{sec:broad_fe-line_model}).  In addition, two-parameter joint-confidence contours were created for the three lines showing line energy versus line flux.  This produced three sets of closed contours across the Fe\,K band generated at 68, 90, 99 and 99.9\,per cent confidence.  These are shown in Figure~\ref{fig:xis03_fe_three_gaussian_contours} and suggest that three significantly detected narrow lines may be a feasible desciption of the {\sl Suzaku} XIS data.

A more physically motivated approach was made by modelling the peaks blueward of 6.4\,keV with a model comprising the emission from optically-thin, photoionized gas computed using \textsc{xstar} (see \citealt{Kallman04}).  The \textsc{pexmon} component was re-included which models the peak at $\sim$6.4\,keV.  The \textsc{xstar} code calculates the physical conditions assuming a slab of photoionized gas illuminated by a central point-source of ionizing radiation and is largely parametrized by its column density, $N_{\rm H}$, and ionization parameter, $\xi$.  

We assume a turbulence velocity of $v_{\rm turb} = 200$\,km\,s$^{-1}$ and an ionizing continuum taking the form of a power law with $\Gamma = 2.2$ (a typical value for this source).  The ionizing luminosity was assumed to be $L_{\rm ion} = 3 \times 10^{43}$\,erg\,s$^{-1}$ and is calculated by extrapolating the continuum model from 1--1\,000\,Rydberg.  Fixing the \textsc{xstar} zone at the redshift of the source, it has a best-fitting ionization parameter of log\,$\xi \sim 3.5$ and contributes emission lines at $\sim$6.7 and $\sim$6.95\,keV due to Fe\,\textsc{xxv} and Fe\,\textsc{xxvi}\,Ly$\alpha$ respectively.  However, in the case of an emission grid there is a strong degree of degeneracy between its column density and normalization.  For a uniform, Compton-thin shell, the normalization of the emission component is defined within \textsc{xstar} as:

\begin{equation} \label{eq:xstar_normalization} k = f_{\rm cov} \frac{L_{\rm ion}}{D^{2}}, \end{equation}

where $f_{\rm cov}$ is the covering fraction of the material, $L_{\rm ion}$ is the ionizing luminosity in units $10^{38}$\,erg\,s$^{-1}$ and $D$ is the distance to the source in kpc.  Physically, the \textsc{xstar} normalization is proportional to the product of the observed X-ray flux and the covering fraction of the photoionized gas.  Assuming $f_{\rm cov} = 1$, the normalization can then be calculated to be $k = 4 \times 10^{33}$\,erg\,s$^{-1}$\,kpc$^{-2}$ for the case of IRAS 18325-5926.  Fixing the normalization of the \textsc{xstar} zone to this value in the model and re-fitting then provides a lower limit on the column density assuming a fully-covering spherical shell of gas.  In this case, the 90\,per cent confidence lower limit is $N_{\rm H} > 5.8 \times 10^{23}$\,cm$^{-2}$.  Note that the column density drops to $N_{\rm H} > 4.2 \times 10^{23}$\,cm$^{-2}$ if $z$ (and hence the outflow velocity, $v_{\rm out}$) is allowed to vary.  The confidence range on $z$ provides a 90\,per cent confidence interval for $v_{\rm out}$ of 100--3\,400\,km\,s$^{-1}$ and corresponds to $\Delta \chi^{2} \sim 3$ between being free and frozen at 0.

\begin{figure}
\begin{center}
\rotatebox{-90}{\includegraphics[width=6cm]{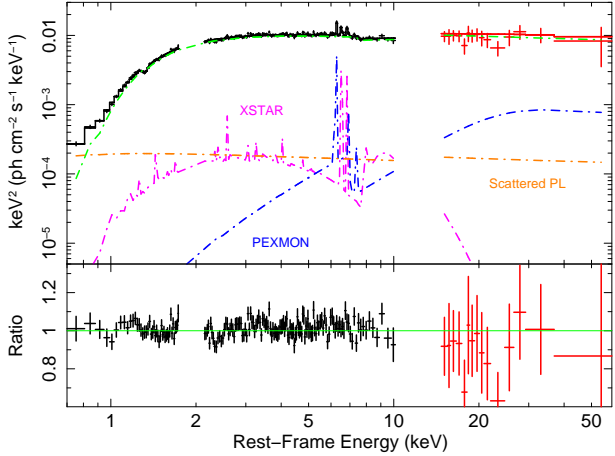}}
\end{center}
\caption{The fit to the broad-band {\sl Suzaku} spectrum of IRAS 18325-5926 when modelled with a photoionized emission component, as described in Section~\ref{sec:narrow_fe_lines}.  The XIS and HXD-PIN data are shown in black and red respectively.  The additive model components are the primary absorbed power law (green), the scattered power law (orange), a neutral reflector (navy) and a photoionized emission grid (magenta).  The bottom panel shows the data-to-model residuals.}
\label{fig:xis03_pin_tbabs_po_po_pexmon_xstar_gauss_edge_eeuf_ratio}
\end{figure}

\begin{figure}
\begin{center}
\rotatebox{-90}{\includegraphics[width=6cm]{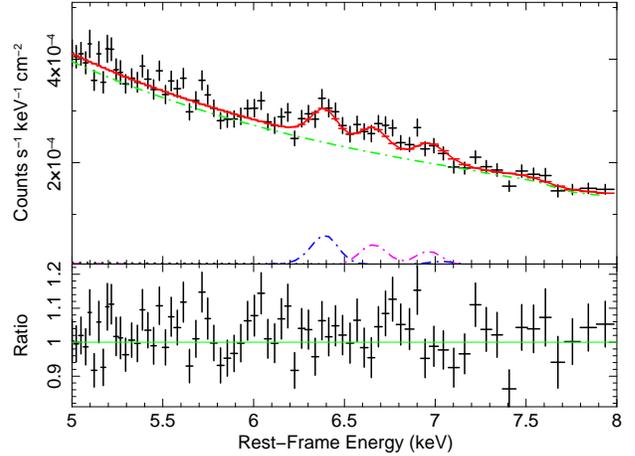}}
\end{center}
\caption{The fit to the {\sl Suzaku} spectrum of IRAS 18325-5926 from 5--8\,keV when the Fe\,K band is modelled with a photoionized emission component, as described in Section~\ref{sec:narrow_fe_lines}.  The emission from the \textsc{pexmon} neutral reflector is shown in navy while the photoionized emission grid is shown in magenta.  The primary power-law component is shown in green while the overall model is shown in red.  The bottom panel shows the data-to-model residuals.}
\label{fig:xis03_pin_pexmon_xstar_fe_data_ratio}
\end{figure}

Finally, as in Section~\ref{sec:broad_fe-line_model}, a sweep across the 7--8\,keV band with the \textsc{edge} model results in best-fitting parameters for the energy of $E_{\rm edge} = 7.49^{+0.14}_{-0.29}$\,keV and the optical depth of $\tau = 0.13 \pm 0.05$ ($\Delta \chi^{2} = 22$).  This model results in a fit statistic to $\chi^{2}$/d.o.f. $= 1508 / 1421$ by accounting for some of the residuals bluewards of 6.4\,keV.  This is a slight improvement over the ionized-reflection model described above and is shown in Figure~\ref{fig:xis03_pin_tbabs_po_po_pexmon_xstar_gauss_edge_eeuf_ratio}.  However, the \textsc{xstar} component still leaves some excess emission unmodelled, particularly around 6.7--6.9\,keV where the residuals don't quite match the Fe\,\textsc{xxv} and Fe\,\textsc{xxvi} peaks in the \textsc{xstar} emission grid (see Figure~\ref{fig:xis03_pin_pexmon_xstar_fe_data_ratio})\footnote{We note that accounting for the peaks blueward of 6.4\,keV with an alternative photoionized emission grid of higher turbulence velocity ($v_{\rm turb} = 1\,000$--$3\,000$\,km\,s$^{-1}$) does not significantly improve the fit ($\Delta \chi^{2} = 1$), requiring higher best-fitting values for the ionization parameter and column density of log\,$\xi \sim 4$ and $N_{\rm H} \sim 10^{24}$\,cm$^{-2}$ to match the observed equivalent widths.}.  The fit statistic achieved over the 5.5--7.5\,keV band is $\chi^{2}$/d.o.f. $= 282 / 223$ (see Table~\ref{tab:model_parameters}).

\begin{table*}
\centering
\begin{tabular}{l c c c c c}
\toprule
\multirow{2}{*}{Model Component} & \multirow{2}{*}{Parameter} & \multirow{2}{*}{Value} & \multirow{2}{*}{$\Delta \chi^{2}$} & $\chi^{2} / {\rm d.o.f.}$ & $\chi^{2} / {\rm d.o.f.}$ \\ 
& & & & (0.7--50\,keV) & (5.5--7.5\,keV) \\ [0.5ex]
\midrule
Power law$^{a}_{\rm int.}$ & $\Gamma$ & $2.19^{+0.02}_{-0.02}$ & & $1564 / 1425$ & $310 / 227$\\ 
& Norm & $1.32^{+0.04}_{-0.04} \times 10^{-2}$ & & & \\
Power law$^{a}_{\rm scatt.}$ & $\Gamma$ & $2.19^{+0.02}_{-0.02}$ & 106 & & \\ 
& Norm & $2.63^{+0.39}_{-0.39} \times 10^{-4}$ & & & \\
\textsc{tbabs}$^{b}$ & $N_{\rm H}$ & $1.60^{+0.04}_{-0.04} \times 10^{22}$ & \\
\textsc{pexmon}$^{c}$ & $R$ & $0.36^{+0.11}_{-0.10}$ & 41 & & \\
\midrule
\textsc{pexmon}$^{c}$ & $R$ & $0.15^{+0.11}_{-0.12}$ & & $1513 / 1421$ & $291 / 223$ \\
\textsc{rdblur} $\times$ \textsc{reflionx}$^{d}$ & $\xi$ & $1\,000^{+1\,100}_{-300}$ & 13 & & \\
& Norm & $1.68^{+0.10}_{-0.10} \times 10^{-8}$ & & & \\
\textsc{edge}$^{e}$ & $E_{\rm c}$ & $7.16^{+0.11}_{-0.05}$ & 26 & & \\
& $\tau$ & $0.13^{+0.04}_{-0.04}$ & & & \\
\midrule
\textsc{pexmon}$^{c}$ & $R$ & $0.35^{+0.10}_{-0.11}$ & & $1508 / 1421$ & $282 / 223$ \\
\textsc{xstar}$^{f}$ & $N_{\rm H}$ & $> 5.8 \times 10^{23}$ & 21 & & \\
& log\,$\xi$ & $3.47^{+0.02}_{-0.04}$ & & & \\
\textsc{edge}$^{e}$ & $E_{\rm c}$ & $7.49^{+0.14}_{-0.29}$ & 22 & & \\
& $\tau$ & $0.13^{+0.05}_{-0.05}$ & & & \\
\bottomrule
\end{tabular}
\caption{The best-fitting parameters of the various models applied to the Fe\,K band described in Sections~\ref{sec:broad_fe-line_model} and~\ref{sec:narrow_fe_lines}.  $^{a}\Gamma$, photon index; normalization in units photons\,cm$^{-2}$\,s$^{-1}$.  $^{b}N_{\rm H}$, column density in units cm$^{-2}$.  $^{c}R$, reflection scaling fraction; unitless.  $^{d}\xi$, ionization parameter in units erg\,cm\,s$^{-1}$; normalization in units photons\,cm$^{-2}$\,s$^{-1}$.  $^{e}$$E_{\rm c}$, centroid energy in units keV; $\tau$, optical depth.  $^{f}N_{\rm H}$, column density in units cm$^{-2}$; $\xi$, ionization parameter in units erg\,cm\,s$^{-1}$.  A covering fraction of $f_{\rm cov} = 1$ is assumed for the \textsc{xstar} emission grid.  All $\Delta \chi^{2}$ values correspond to the change in the fit statistic upon removal of the component from the overall model.}
\label{tab:model_parameters}
\end{table*}

\subsection{X-ray spectral variability} \label{sec:X-ray_spectral_variability}

\begin{figure}
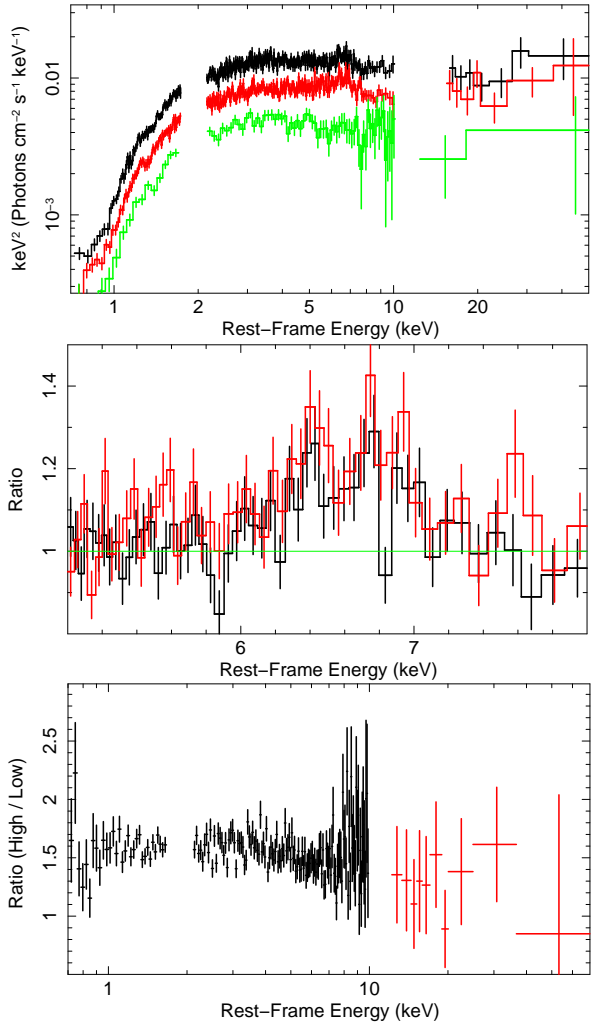

\begin{center}
\rotatebox{-90}{\includegraphics[width=4.5cm]{xis03_pin_high_low_difference.ps}}
\rotatebox{-90}{\includegraphics[width=4.5cm]{xis03_high_low_fe_ratio.ps}}
\rotatebox{-90}{\includegraphics[width=4.5cm]{xis03_pin_ratio.ps}}
\end{center}
\caption{Upper panel: High-flux (black) and low-flux (red) FI-XIS and HXD-PIN spectra from 0.7--50\,keV.  The difference spectrum is shown in green and the FI-XIS data are binned relative to the half-width at half-maximum (HWHM) of the detector resolution.  Middle panel: The ratio of the residuals of the FI-XIS data from 5--8\,keV compared to a power law absorbed by neutral material.  The high- and low-flux spectra are shown in black and red respectively and the data are binned up by a factor of 4 for clarity.  Lower panel: The ratio spectrum between the periods of high and low flux of IRAS 18325-5926.  The ratios of the FI-XIS and HXD-PIN data are shown in black and red respectively.}
\label{fig:xis03_pin_high_low_diff_ratio}
\end{figure}

The XIS light curve displayed in Figure~\ref{fig:lightcurves} shows that IRAS 18325-5926 displays significant X-ray variability at energies $< 10$\,keV on timescales of tens of ks.  In an attempt to quantify this high- and low-flux FI-XIS spectra were created by using only the counts from the first and second halves of the observation respectively (i.e. splitting after the 13th bin in Figure~\ref{fig:lightcurves}).  The same time selection was also applied to the HXD data to generate high- and low-flux PIN spectra.  IRAS 18325-5926 is less variable in the HXD-PIN band and results in 15--50\,keV count rates of $0.042 \pm 0.004$ and $0.034 \pm 0.003$\,counts\,s$^{-1}$ for the high- and low-flux segments respectively.  The ``fluxed'' high- and low-flux broad-band X-ray spectra are shown in Figure~\ref{fig:xis03_pin_high_low_diff_ratio} (upper panel).

The Fe\,K band residuals to a simple absorbed power-law continuum fitted between 2.5--5.5 and 7.5--10\,keV are shown in Figure~\ref{fig:xis03_pin_high_low_diff_ratio} (middle panel).  Fitting the Fe\,K band with the models described in Sections~\ref{sec:broad_fe-line_model} and~\ref{sec:narrow_fe_lines} results in fits that are consistent with those reported in Table~\ref{tab:model_parameters}.

The high- and low-flux broad-band spectra were fitted with the models described in Sections~\ref{sec:broad_fe-line_model} and~\ref{sec:narrow_fe_lines}.  Consistent with \citet{Tripathi13}, we find that simply allowing the normalization of the primary (intrinsic) power-law component to vary while keeping all other free parameters tied together between the two datasets accounts for the majority of the observed spectral variability well.  The normalization varies by a factor of $\sim$1.5 while the ratio of scattered to intrinsic flux varies from $\sim$0.014--0.023.  The resultant fit statistic (in the case of the model described in Section~\ref{sec:narrow_fe_lines}) from 0.7--50\,keV is $\chi^{2}$/d.o.f. = $2061/1925$ ($\chi^{2}$/d.o.f. = $1312/1300$ from 2.5--50\,keV).  We note that the fit statistic can be improved further ($\Delta \chi^{2} \approx 10$) by also allowing the normalization of the scattered power-law component to vary with respect to the models described in Sections~\ref{sec:broad_fe-line_model} and~\ref{sec:narrow_fe_lines}.  This then results in the ratio of scattered to intrinsic flux remaining at a constant value of $\sim$0.02.  However, a changing flux of the presumably distant, scattered emission component would be difficult to explain on such a short timescale and, additionally, an identical improvement can be found to the fit by allowing the photon index of the power-law to vary ($\Delta \Gamma = 0.04$) with no requirement for both $\Gamma$ and the normalization of the scattered power-law to simultaneously vary.

We have also examined and fitted the difference and ratio spectra (e.g. \citealt{FabianVaughan03}) to search for indications of more complex spectral variability.  The results were consistent with those described above.  Difference spectra were created by subtracting the low-flux source and background spectra from their respective high-flux spectra, propagating the errors during each mathematical operation.  If the Fe\,K emission component does arise from the inner regions of an ionized accretion disc, the broad component of emission may be expected to vary in phase with the continuum.  Consequently, one may anticipate some of the broad emission to show up in the difference spectrum since any constant components should have been subtracted away.  Using a \textsc{diskline} component \citep{Fabian89} to fit the broad Fe line in the difference spectrum (Figure~\ref{fig:xis03_pin_high_low_diff_ratio}; upper panel), we found that this component is not required by the data and gave an upper limit on the flux of $8.4 \times 10^{-6}$\,photons\,cm$^{-2}$\,s$^{-1}$, which is $< 20$ per cent of the line flux in the time-averaged spectrum.  By contrast, the 2--10\,keV flux changes by $\sim$50 per cent.  Likewise the flux from 7--10\,keV (the important energy band for driving Fe\,K fluorescence) also changes by $\sim$50 per cent.

Ratio spectra were created by dividing the background-subtracted high-flux spectrum by the background-subtracted low-flux spectrum, having applied identical binning criteria.  The high/low ratio spectrum (Figure~\ref{fig:xis03_pin_high_low_diff_ratio}; lower panel) is observed to be generally flat with a uniform difference in the count rate of a factor of $\sim$1.5.  Again. this is consistent with a scenario in which the variability can be largely explained in terms of a single component varying in normalization.

\subsection{Previous observations} \label{sec:previous_observations}

\subsubsection{{\sl XMM-Newton}} \label{sec:xmm-newton}

The shape of the 2001 broad-band {\sl XMM-Newton} MOS continuum can be largely reproduced by the model described in Section~\ref{sec:X-ray_continuum} but with a higher ratio of scattered-to-primary continuum flux of $\sim$0.03, consistent with \citet{Tripathi13}.  The photon index of the power-law continuum is harder than in the {\sl Suzaku} spectrum ($\Gamma = 1.94 \pm 0.02$ versus $\Gamma = 2.19 \pm 0.02$) while the source flux is nearly a factor of 2 lower suggestive of steeper-when-brighter continuum behaviour.  Consistent with \citet{Tripathi13}, the normalization of the scattered power law varies by $\sim$50 per cent, indicating that the scattered component may respond to the primary continuum on long timescales.

In the soft X-ray band, several dips can also be observed in the MOS spectrum, none of which has been reported before in analyses of these {\sl XMM-Newton} data.  If associated with narrow absorption lines the most notable of these features occur at the following rest-frame energies (when modelled with Gaussians with $\sigma = 0$\,eV): $E_{\rm c} = 0.97 \pm 0.01$\,keV (EW $= 15 \pm 3$\,eV; $\Delta \chi^{2} =40$); $E_{\rm c} = 1.07^{+0.02}_{-0.01}$\,keV (EW $= 8 \pm 4$\,eV; $\Delta \chi^{2} = 18$); $E_{\rm c} = 1.21 \pm 0.02$\,keV (EW $= 6 \pm 3$\,eV; $\Delta \chi^{2} = 14$) and $E_{\rm c} = 1.34^{+0.01}_{-0.02}$\,keV (EW $= 10 \pm 2$\,eV; $\Delta \chi^{2} = 47$).  We are only able to confidently identify the feature at 1.34\,keV which is most likely due to Mg\,\textsc{xi}, consistent with the {\sl Suzaku} XIS (Section~\ref{sec:X-ray_continuum}) and {\sl Chandra} HETG spectra \citep{Mocz11}.  The feature at 0.97\,keV lies close to the expected energy of the Ne\,\textsc{x} Ly$\alpha$ transition (1.02\,keV) but may also be blended with various inner-shell transitions from Fe\,\textsc{xvii}.  Finally, the lines at 1.07\,keV and 1.20\,keV may also be associated with Ne with the features arising from the Ne\,\textsc{ix} $1s$--$3p$ transition (1.07\,keV) and the Ne\,\textsc{x} Ly$\beta$ transition (1.21\,keV) respectively.

Two additional narrow features are also seen as dips at observed energies of $E_{\rm c} = 0.58^{+0.02}_{-0.01}$\,keV (EW $< 30$\,eV; $\Delta \chi^{2} = 10$) and $E_{\rm c} = 1.84 \pm 0.02$\,keV (EW $= 11^{+4}_{-3}$\,eV; $\Delta \chi^{2} = 26$).  The former roughly coincides with the O\,\textsc{i} edge in the EPIC-MOS detectors at 0.54\,keV where the calibration due to the filter transmission may have degraded by up to $\simeq 10$\,per cent in the case of the MOS\,1 camera\footnote{http://xmm2.esac.esa.int/external/xmm\_sw\_cal/calib/rel\_notes/XMM-CCF-REL-273}.  Interestingly this feature is only apparent in the MOS\,2 camera.  As such, and in the absence of any useful RGS data, we cannot say with confidence whether this feature is real or if its origin lies in the instrumental response.  Regarding the latter feature, its observed energy coincides with that of the Si\,K edge at 1.84\,keV and so likewise we are unable to confidently assess the reality of the feature.

\begin{figure}
\begin{center}
\rotatebox{-90}{\includegraphics[width=4.5cm]{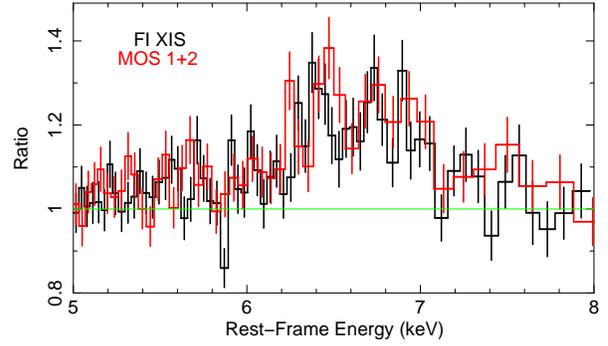}}
\end{center}
\caption{The combined {\sl XMM-Newton} MOS spectrum overlayed on the {\sl Suzaku} FI-XIS spectrum plotted as a ratio to an absorbed power-law continuum in the Fe\,K band.}
\label{fig:xis03_mos12_fe_ratio}
\end{figure}

Figure~\ref{fig:xis03_mos12_fe_ratio} shows the combined {\sl XMM-Newton} MOS spectrum as a ratio to an absorbed power-law continuum overlayed on the combined {\sl Suzaku} FI-XIS spectrum in the Fe\,K band.  The Fe-line complex is largely visually consistent between the two observations despite them being made 6 years apart.  However, one difference is that, unlike the XIS spectrum, the MOS profile has no obvious peak at 6.4\,keV, instead peaking at $\sim$6.5\,keV.  Consequently, a neutral reflection component, such as the \textsc{pexmon}, slightly over-predicts the flux at 6.4\,keV.  Including an \textsc{xstar} photoionization emission component, as in Section~\ref{sec:narrow_fe_lines}, to model the higher-energy peaks results in best-fitting values consistent with those reported in Table~\ref{tab:model_parameters} (log\,$\xi = 3.57^{+0.28}_{-0.10}$; $N_{\rm H} > 3.2 \times 10^{23}$\,cm$^{-2}$).  However, similar to the XIS spectrum, the photoionization grid leaves some excess emission at $\sim$6.5 and $\sim$6.8\,keV unmodelled.  Again, increasing the turbulence velocity of the emission component does not significantly improve the fit.  The MOS spectrum also reveals the presence of an absorption trough peaking at $\sim$8.5\,keV; an absorption edge at $E_{\rm edge} = 7.77^{+0.27}_{-0.21}$\,keV with an optical depth of $\tau = 0.19 \pm 0.06$ can account for this well, improving the fit by $\Delta \chi^{2} = 30$.  This is consistent with the {\sl Suzaku} XIS spectrum (Section~\ref{sec:narrow_fe_lines}).  The resultant fit statistic over the 0.2--10\,keV energy range for this model is $\chi^{2}$/d.o.f. $= 2401 / 2355$ ($\chi^{2}$/d.o.f. $= 346 / 303$ from 5.5--7.5\,keV).

A better fit can be obtained using the blurred reflection model described in Section~\ref{sec:broad_fe-line_model} (i.e. \textsc{reflionx} convolved with \textsc{rdblur}).  This requires the reflector to be ionized with $\xi = 800^{+800}_{-500}$\,erg\,cm\,s$^{-1}$ with an inclination angle of $\theta_{\rm obs} = 31^{+16}_{-8}$\,deg.  However, the emissivity index and inner radius again remain unconstrained and so are fixed at $q = 2$ and $r_{\rm in} = 6$\,$r_{\rm g}$ respectively.  We note that the \textsc{pexmon} neutral-reflection component does not significantly improve the fit ($R = 0.16^{+0.15}_{-0.12}$).  However, by including a narrow ($\sigma = 10$\,eV) Gaussian fixed at 6.4\,keV, the constraint on the flux of the near-neutral narrow core is $F_{\rm K\alpha} = 4.0^{+2.0}_{-2.3} \times 10^{-6}$\,photons\,cm$^{-2}$\,s$^{-1}$ which is consistent within the uncertainties with the flux of the same component in the equivalent {\sl Suzaku} model ($F_{\rm K\alpha} = 7.4^{+3.2}_{-3.1} \times 10^{-6}$\,photons\,cm$^{-2}$\,s$^{-1}$)\footnote{Note that performing the same test with the `narrow' \textsc{xstar} model also shows that the flux of the narrow Fe\,K$\alpha$ core is consistent between {\sl Suzaku} and {\sl XMM-Newton} within the uncertainties: $F_{\rm K\alpha} = 1.2 \pm 0.3 \times 10^{-5}$ ({\sl Suzaku}); $F_{\rm K\alpha} = 7.3^{+3.1}_{-2.2} \times 10^{-6}$\,photons\,cm$^{-2}$\,s$^{-1}$ ({\sl XMM-Newton}).}.  This model results in a fit statistic of $\chi^{2}$/d.o.f. $= 2395 / 2354$ ($\chi^{2}$/d.o.f. $= 341 / 302$ from 5.5--7.5\,keV).  The best-fitting parameters are summarized in Table~\ref{tab:model_parameters_xmm_asca}.

\subsubsection{{\sl ASCA}} \label{sec:asca}

Similar to the {\sl XMM-Newton} MOS spectra, the {\sl Suzaku} model described in Section~\ref{sec:X-ray_continuum} provides a reasonable description of the shape of the broad-band properties of the {\sl ASCA} spectrum with a ratio of scattered-to-primary continuum flux of $\sim$0.02, consistent with the {\sl Suzaku} data.  The flux of the source falls in between the {\sl Suzaku} and {\sl XMM-Newton} observations and, likewise, so does the photon index of the power-law continuum ($\Gamma = 2.05 \pm 0.03$; compare with Section~\ref{sec:xmm-newton}).

Attempting to account for the Fe\,K band emission with photoionized gas (as in Section~\ref{sec:narrow_fe_lines}) returns best-fitting values consistent with those obtained with {\sl Suzaku} and {\sl XMM-Newton} but with larger uncertainties (log\,$\xi = 3.64^{+0.22}_{-1.07}$; $N_{\rm H} > 1.6 \times 10^{23}$).  The reflection scaling factor of the \textsc{pexmon} component becomes $R = 0.42^{+0.15}_{-0.18}$, consistent with {\sl Suzaku}.  The \textsc{xstar} grid improves the fit by $\Delta \chi^{2} = 9$ and results in a broad-band fit statistic from 0.7--10\,keV of $\chi^{2}$/d.o.f. $= 502/464$ ($\chi^{2}$/d.o.f. $= 291/260$ from 2.5--10\,keV and $\chi^{2}$/d.o.f. $= 86/61$ from 5.5--7.5\,keV).

Like with {\sl Suzaku} and {\sl XMM-Newton}, attempting to model the {\sl ASCA} spectra instead with a blurred reflection component (Section~\ref{sec:broad_fe-line_model}) again requires an origin in an ionized disc with $\xi > 2 \times 10^{3}$\,erg\,cm\,s$^{-1}$.  However, the model is not well constrained, requiring parameters to be fixed at $r_{\rm in} = 6$\,$r_{\rm g}$, $q = 2$ and inclination angle $\theta_{\rm obs} = 60$\,deg.  As in Section~\ref{sec:xmm-newton}, the \textsc{pexmon} component does not formally improve the fit ($R < 0.40$) but a better overall fit can be obtained with this blurred reflection model with $\chi^{2}$/d.o.f. $= 496/464$ ($\chi^{2}$/d.o.f. $= 282/260$ from 2.5--10\,keV and $\chi^{2}$/d.o.f. $= 77/61$ from 5.5--7.5\,keV; see Table~\ref{tab:model_parameters_xmm_asca}).  However, like with {\sl Suzaku} and {\sl XMM-Newton}, none of the models appears to offer a complete description of the emission complex.

\begin{table*}
\centering
\begin{tabular}{l c c c c c c c}
\toprule
& & \multicolumn{3}{c}{{\sl XMM-Newton} MOS} & \multicolumn{3}{c}{{\sl ASCA}} \\
\multirow{2}{*}{Model Component} & \multirow{2}{*}{Parameter} & \multirow{2}{*}{Value} & $\chi^{2} / {\rm d.o.f.}$ & $\chi^{2} / {\rm d.o.f.}$ & \multirow{2}{*}{Value} & $\chi^{2} / {\rm d.o.f.}$ & $\chi^{2} / {\rm d.o.f.}$ \\ 
& & & (0.2--10\,keV) & (5.5--7.5\,keV) & & (0.7--10\,keV) & (5.5--7.5\,keV) \\ [0.5ex]
\midrule
Power law$^{a}_{\rm int.}$ & $\Gamma$ & $1.94^{+0.02}_{-0.02}$ & $2454 / 2359$ & $370 / 307$ & $2.05^{+0.03}_{-0.03}$ & $542 / 468$ & $103 / 65$ \\ 
& Norm & $5.31^{+0.13}_{-0.17} \times 10^{-3}$ & & & $7.87^{+0.30}_{-0.28} \times 10^{-3}$ & & \\
Power law$^{a}_{\rm scatt.}$ & $\Gamma$ & $1.94^{+0.02}_{-0.02}$ & & & $2.05^{+0.03}_{-0.03}$ & & \\ 
& Norm & $1.41^{+0.06}_{-0.07} \times 10^{-4}$ & & & $1.80^{+0.35}_{-0.35} \times 10^{-4}$ \\
\textsc{tbabs}$^{b}$ & $N_{\rm H}$ & $1.25^{+0.03}_{-0.03} \times 10^{22}$ & & & $1.68^{+0.05}_{-0.05} \times 10^{22}$ & & \\
\textsc{pexmon}$^{c}$ & $R$ & $0.39^{+0.10}_{-0.11}$ & & & $0.56^{+0.16}_{-0.17}$ & & \\
\midrule
\textsc{pexmon}$^{c}$ & $R$ & $0.16^{+0.15}_{-0.12}$ & $2395 / 2354$ & $341 / 302$ & $< 0.40$ & $496 / 464$ & $77 / 61$ \\
\textsc{rdblur} $\times$ \textsc{reflionx}$^{d}$ & $\xi$ & $800^{+800}_{-500}$ & & & $> 2 \times 10^{3}$ & & \\
& $\theta_{\rm obs}$ & $31^{+16}_{-8}$ & & & $60$ & & \\
& Norm & $1.63^{+1.41}_{-0.41} \times 10^{-8}$ & & & $6.65^{+6.59}_{-5.48} \times 10^{-8}$ & & \\
\textsc{edge}$^{e}$ & $E_{\rm c}$ & $7.85^{+0.27}_{-0.26}$ & & & $8.54^{+0.50}_{-0.43}$ & & \\
& $\tau$ & $0.19^{+0.07}_{-0.06}$ & & & $0.15^{+0.18}_{-0.11}$ & & \\
\midrule
\textsc{pexmon}$^{c}$ & $R$ & $0.34^{+0.11}_{-0.10}$ & $2401 / 2355$ & $346 / 303$ & $0.42^{+0.15}_{-0.18}$ & $502 / 464$ & $86 / 61$ \\
\textsc{xstar}$^{f}$ & $N_{\rm H}$ & $> 3.2 \times 10^{23}$ & & & $> 1.6 \times 10^{23}$ & & \\
& log\,$\xi$ & $3.57^{+0.28}_{-0.10}$ & & & $3.64^{+0.22}_{-1.07}$ & & \\
\textsc{edge}$^{e}$ & $E_{\rm c}$ & $7.77^{+0.27}_{-0.21}$ & & & $8.51^{+0.30}_{-0.26}$ & & \\
& $\tau$ & $0.19^{+0.06}_{-0.06}$ & & & $0.32^{+0.13}_{-0.11}$ & & \\
\bottomrule
\end{tabular}
\caption{The best-fitting parameters of the various models applied to the {\sl XMM-Newton} and {\sl ASCA} datasets described in Sections~\ref{sec:xmm-newton} and~\ref{sec:asca}.  $^{a}\Gamma$, photon index; normalization in units photons\,cm$^{-2}$\,s$^{-1}$.  $^{b}N_{\rm H}$, column density in units cm$^{-2}$.  $^{c}R$, reflection scaling fraction; unitless.  $^{d}\xi$, ionization parameter in units erg\,cm\,s$^{-1}$; $\theta_{\rm obs}$, inclination angle in units of degrees; normalization in units photons\,cm$^{-2}$\,s$^{-1}$.  $^{e}$$E_{\rm c}$, centroid energy in units keV; $\tau$, optical depth.  $^{f}N_{\rm H}$, column density in units cm$^{-2}$; $\xi$, ionization parameter in units erg\,cm\,s$^{-1}$.  A covering fraction of $f_{\rm cov} = 1$ is assumed for the \textsc{xstar} emission grid.}
\label{tab:model_parameters_xmm_asca}
\end{table*}

\section{Discussion \& Conclusions} \label{sec:discussion_and_conclusions}

In this paper we have examined the {\sl Suzaku} XIS+HXD, {\sl XMM-Newton} EPIC-MOS and {\sl ASCA} SIS+GIS data of IRAS 18325-5926.  These are the highest $S$/$N$ CCD-resolution X-ray spectra of this Seyfert 2 galaxy.  We find that the broad-band properties are largely consistent with those reported by previous authors with the X-ray spectrum dominated by a primary power-law continuum ($\Gamma \sim 2.2$) modified by a moderate column of cold absorbing gas ($N_{\rm H} \sim 10^{22}$\,cm$^{-2}$).  A weak component of scattered emission is apparent in the soft X-ray band most likely arising in distant material.  We also find a complex Fe-line emission profile; the shape of the residuals against a simple continuum model appear broad but show tentative evidence of comprising multiple lines.  In particular, the structure of the emission profile in the highest-resolution data (i.e the {\sl Suzaku} XIS data) appears to display peaks arising at $\sim$6.4--6.5, $\sim$6.7--6.8 and $\sim$6.9\,keV.  Previous studies of this source have concentrated on fitting stand-alone broad components to the Fe line but here we explore a wider range of models and more physical broad-band models are tested.

In Sections~\ref{sec:fe-line_complex}--\ref{sec:previous_observations} we fit the data using a range of models that explore different origins for the Fe\,K emission features.  None of the models provides a particularly good fit from 5.5--7.5\,keV.  The {\sl Suzaku} XIS spectrum appears to be better fitted by three narrow emission lines (neutral reflection plus photoionized gas) whereas a blurred ionized reflector offers a better fit to the {\sl XMM-Newton} MOS spectrum, largely on the grounds that the 6.4\,keV peak in the XIS spectrum (associated with neutral reflection) is weaker in the EPIC-MOS data with additional residuals at $\sim$6.5\,keV.  The physical validity of each model is discussed below.

In Section~\ref{sec:broad_fe-line_model}, the {\sl Suzaku} spectrum was fitted with a broad-band reflection model whereby the \textsc{reflionx} code of \citet{RossFabian05} was convolved with the \textsc{rdblur} kernel to produce a blurred reflection spectrum.  Assuming this model, the best-fitting parameters suggest an origin in an ionized accretion disc for the Fe-line complex; a scenario which may be supported by the evidence for a weak absorption edge $> 7.11$\,keV.  Furthermore, the weak Compton-reflection hump observed in the broad-band {\sl Suzaku} XIS+HXD spectrum suggests that it is unlikely that the Fe-line profile is dominated by reflection from cold, neutral material.  Such an origin in ionized material for the Fe line was favoured by several previous authors (e.g. \citealt{Iwasawa96,Iwasawa04,Mocz11}).  If such a scenario were true, it may be conceivable that we are afforded a view of the nuclear region in what is otherwise a type-2 system due to either a global covering of patchy gas (e.g. \citealt{Iwasawa95}) or a unique viewing angle to the source whereby our sightline just grazes the putative molecular torus (e.g. \citealt{Mocz11}).

The data appear unable to precisely constrain the innermost radius of emission - however, the degree of blurring required to broaden the Fe\,K emission features appears consistent with an origin $\sim$6\,$r_{\rm g}$ from the black hole (i.e. the ISCO in the case of a Schwarzschild geometry).  Should the bulk of the emission originate from close to the black hole, one may reasonably expect the Fe\,K emission profile to vary in phase with the illuminating continuum.  The Fe\,K emission profile was shown to remain largely constant in flux despite changes in the 7--10\,keV continuum flux of $\sim$50 per cent (see Section~\ref{sec:X-ray_spectral_variability} and Figure~\ref{fig:xis03_pin_high_low_diff_ratio}; middle panel).  We note that Fe-line reverberation is a hotly debated topic (e.g. \citealt{Reynolds99}; \citealt{BallantyneTurnerYoung05}; \citealt{Fabian09}) and the detailed physics of the disc reprocessing may result in a more complex connection between the observed line and continuum fluxes.

An alternative model was outlined in Section~\ref{sec:narrow_fe_lines} whereby the Fe-line profile was instead explained in terms of a neutral reflector peaking at $\sim$6.4\,keV\footnote{We note that \citet{Mocz11} do test for the presence of a narrow core of emission from Fe and state that the presence of one is unrequired - however, this may be due to the lower photon statistics in their {\sl Chandra} HETG spectrum.  Indeed there are $> 20$ times fewer counts in the {\sl Chandra} HEG spectrum compared with the combined FI-XIS {\sl Suzaku} spectrum.} plus additional narrow, ionized emission lines peaking at 6.72\,keV and 6.94\,keV.  The peak at 6.4\,keV is statistically required in the {\sl Suzaku} spectrum and its flux is consistent within the uncertainties between observations.  When the absorber and reflector are self-consistently modelled with the \textsc{mytorus} model (Section~\ref{sec:mytorus_model}), the flux at 6.4\,keV is slightly underpredicted unless the normalizations of the emission components are allowed to be larger than that of the absorber.  Such a scenario could arise if Fe is mildly overabundant with respect to solar values or if there is an additional source of near-neutral Fe emission.  Alternatively, the column density of the scattering / line-producing material may be higher than that of the line-of-sight absorber.  This could be consistent with a picture where our sightline grazes lower-column material at the edge of the torus while the majority of the scattered emission originates from higher-density clumps away from our line of sight.

The highly-ionized lines at $\sim$6.7\,keV and $\sim$6.95\,keV may then originate in photoionized gas.  Accounting for their presence with \textsc{xstar} offers a slight improvement over the ionized-reflection model when fitted to the broad-band {\sl Suzaku} spectrum and requires the gas to be highly ionized (log\,$\xi \sim 3.5$) with the narrow lines arising from the $1s$--$2p$ resonance transition from Fe\,\textsc{xxv} and the Ly$\alpha$ transition from Fe\,\textsc{xxvi}.  Their equivalent widths are found to be EW $=38^{+19}_{-13}$\,eV and EW $=33^{+18}_{-15}$\,eV respectively.  \citet{BianchiMatt02} calculate the predicted equivalent widths of Fe\,\textsc{xxv} and Fe\,\textsc{xxvi} emission lines from photoionized, optically-thin gas against the total continuum.  For the case where log\,$\xi \sim 3.5$ and $\Gamma \sim 2.2$ (consistent with the best-fitting values observed here), the predicted equivalent widths of the Fe\,\textsc{xxv} and Fe\,\textsc{xxvi} emission lines are $\sim$40--50\,eV and $\sim$15\,eV respectively.  As such, our observed Fe\,\textsc{xxv} equivalent width is consistent with the predictions of \citet{BianchiMatt02} while the Fe\,\textsc{xxvi} equivalent width may be consistent within the 90\,per cent uncertainties.  The calculations of \citet{BianchiMatt02} assume a column density of $N_{\rm H} = 10^{23}$\,cm$^{-2}$ and so, assuming the predicted equivalent width peaks higher as $N_{\rm H}$ increases, a more consistent equivalent width from Fe\,\textsc{xxvi} emission might be expected for the slightly higher column density required here.

When the model is also applied to the {\sl ASCA} and {\sl XMM-Newton} MOS spectra the best-fitting values are found to be consistent despite the observations taking place over a $\sim$10-year period.  This may suggest an origin in distant material.  One possible origin for such emission could be related to the scattering medium modelled in Section~\ref{sec:X-ray_continuum}.  The nature of such a component was investigated by \citet{Tripathi13} who find that the scatterer must be distant ($3 \times 10^{15} < R < 3 \times 10^{17}$\,cm) and consist of highly-ionized ($\xi > 10^{3}$\,erg\,cm\,s$^{-1}$), high-column-density (a few $10^{23}$\,cm$^{-2}$) gas.  Such values are compatible with the best-fitting values derived from our fits with a photoionized emitter and so it may be conceivable that the distant scattering medium may also produce narrow emission lines from Fe\,\textsc{xxv-xxvi}.  In this scenario the central X-ray source is surrounded by near-neutral material that both absorbs (along the line of sight) the primary X-ray emission and reflects/fluoresces (mostly from material out of the line of sight). Extending beyond this is lower-density photoionized gas that produces a soft continuum by electron-scattering back into our line-of-sight some small fraction of the otherwise absorbed direct continuum plus emission lines and radiation continua from highly-ionized species (e.g. Fe\,\textsc{xxv} and Fe\,\textsc{xxvi}; see \citealt{Kinkhabwala02} for an illustration and further discussion).  We do note that the strength of the scattered-emission component varies by $\sim$50\,per cent between the {\sl XMM-Newton} and {\sl Suzaku} observations (see Section~\ref{sec:xmm-newton}).  As such, it may be reasonable to suspect that the highly-ionized Fe complex may also be variable, although such variations do not appear to be observed between the observations.  Finally, we also find that, when modelling with \textsc{xstar}, the centroid energies of the lines at $\sim$6.7 and $\sim$6.95\,keV don't quite match the energies of the peaks in the XIS spectrum leaving some residuals unmodelled. An increased turbulence velocity does not significantly improve the fit although a slight improvement can be achieved by allowing for a small velocity shift along the line-of-sight - however, this would require the emission to be slightly redshifted.  It therefore may be conceivable that a broader component of Fe emission from an accretion disc is contributing to the overall Fe-line profile and could be responsible for the unmodelled residuals which remain once the narrow features have been modelled.

A final additional source of emitting material may also be related to the outflowing wind detected by \citet{Mocz11} in the {\sl Chandra} HETG spectrum.  Such scenarios have previously been suggested for sources such as the Seyfert 2 NGC 1068 where the radial ionic column densities of the soft X-ray emission features were found to be similar to those found in Seyfert 1 warm absorbers \citep{Kinkhabwala02}.  However, the primary absorber detected by \citet{Mocz11} is found to have column density, $N_{\rm H} \sim 10^{21}$\,cm$^{-2}$ and an ionization parameter, log\,$\xi \sim 2$ (with a possible second higher-ionization absorber; log\,$\xi \sim 2.3-2.6$) and so this medium is unlikely to be responsible for producing photoionized emission from highly-ionized Fe.

To summarize, IRAS 18325-5926 is an interesting type-2 AGN displaying good evidence for a broad emission component in the Fe\,K band which may imply a unique viewing angle to the source, perhaps just grazing the edge of the torus.  As such features are rare in type-2 AGN it is important to assess its true reality.  The best-quality data obtained to date also reveal a peak at 6.4\,keV which is required when modelling the {\sl Suzaku} spectrum and is consistent within the uncertainties between observations.  As such, it may be conceivable that emission from both the inner regions of an ionized accretion disc and from distant, cold material may be contributing to the complex structure of the Fe\,K profile.  However, no model currently offers a complete description of the spectrum since the current data are not of sufficient quality to robustly distinguish between models.  Consequently, better photon statistics (i.e. with the {\sl XMM-Newton} EPIC-pn camera) and/or higher resolution spectra (e.g. with the microcalorimeter on-board the upcoming {\sl Astro-H} mission) are required to disentangle these components and precisely determine the nature of this intriguing Fe-line profile.

\section{Acknowledgements}

This research has made use of the NASA Astronomical Data System (ADS), the NASA Extragalactic Database (NED) and data obtained from the {\sl Suzaku} satellite; a collaborative mission between the space agencies of Japan (JAXA) and the USA (NASA).  This paper is also based on observations obtained with {\sl XMM-Newton}, an ESA science mission with instruments and contributions directly funded by ESA Member States and the USA (NASA).  Additionally, this research has made use of the Tartarus (Version 3.1) database, created by Paul O'Neill and Kirpal Nandra at Imperial College London, and Jane Turner at NASA/GSFC. Tartarus is supported by funding from PPARC, and NASA grants NAG5-7385 and NAG5-7067.  Andrew Lobban would like to acknowledge support from the UK STFC research council.  We also wish to thank our anonymous referee for a careful and thorough review of the draft.

\label{lastpage}

\end{document}